\documentclass[preprintnumbers,amsmath,amssymb]{revtex4}

\usepackage{graphicx}
\usepackage{bm}
\def\be{\begin{equation}}
\def\ee{\end{equation}}
\def\bee{\begin{eqnarray}}
\def\eee{\end{eqnarray}}
\begin{document}

\author{G.Modestino}
\email{modestino@lnf.infn.it}


\affiliation{%
INFN, Laboratori Nazionali di Frascati, via Enrico Fermi 40, I-00044, Frascati (Roma) Italy\\
}%

\date{\today}
\title{Geometrical and physical parameters for electromagnetic relativistic sources.}

\date{\today }

\begin{abstract}

The electromagnetic field is typically measured by the charged particle motion observation. Generally in the experiments, position, velocity and other physical parameters concerning relativistic particle beams, are estimated evaluating the actual distance between the moving source and the detector, at even given time (typically appointing the reference time to the minimum distance between the two points). So to describe faithfully a physical scenario, fixing a spatial reference system is a very  troublesome act as well as evaluating the  geometrical parameters and relative time evolution. In the present note, within a simple experimental context, avoiding as much as possible complex mathematical formalisms, the intensity field evolution is predicted at a given point, using fundamental geometrical assertions. The result is  consistent with the special relativity law. 
\end{abstract}

\maketitle

\section{Introduction}
In several physical context, not only appreciable measurements but also the figurative aspect results crucial for a complete understanding of the real facts and also for a correct communication concerning them. Besides, in several cases,  the complex formalism doesn't allow an exhaustive geometric-spatial representation of the physical environment, and many assumptions are implied for simplifying the description and for making  real  circumstances more adherent to the intuitive sense. The risk is that forgetting the implicit assumptions the measurement result may be distorted or at least ambiguous \cite{fey0}. 
The most typical cases regard physical quantities expressed in terms of four-dimensional vectors, typically in the special relativity, whenever velocities are very close to the light-speed. In the present note, the space-time electromagnetic (EM) field distribution due to moving charges is considered. A large amount of treatments can be found in the fundamental scientific  literature for calculating the complete equations. The general formula can deduced from the Li\'{e}nard-Wiechert potential (in the paper \cite{guido} many references are reported), or equivalently from geometrical space-time considerations, as well explained by R. Feynman\cite{fey1}. We restrict the treatise to the measurement of the progression of the EM field generated from a moving source at constant velocity, observed at a fixed point in the laboratory system. The basic assumption is the effectiveness of a propagation mechanism like the Coulomb law, i.e. the signal dependence on the squared inverse distance from the source. The aim is twofold, to maintain the general validity of the physical law reducing implicit assumptions, and in the mean time, to preserve an intuitive understanding simplifying  the mathematical environment as much as possible.
  \section{Coulomb field evaluation}
\subsection{ The experimental setting}
\label{expe}
 In a surrounding space vacuum where a laboratory system reference $[x,y,z]$ is fixed with origin into ${\bf O}=(0,0,0)$, a point source is moving along the $x$-axis, occupying ${\bf X}(t)=(x(t),0,0)$ at instant $t$.  \\
 The particle moves at constant velocity ${\bf V}=(v_x,0,0)$  on the $x$-axis, such a way always it results 
 \be
 x(t)=v_xt
.
\label{track}
\ee
 
  Along the $y$-line, at distance  $Y$ from the origin $\bf O$, we find an EM detector that records a Coulomb field
 \be
 {\bf E}(t)=k\frac{Q}{R^3(t)}~{{\bf R}(t)}
 \label{efield}
 \ee
 where $k$ is the Coulomb constant, and ${\bf R}(t)$ is a geometrical parameter with components $ [R_x(t),R_y(t),R_z(t)] $.
 As the previous formula shows,  $ {\bf E}(t)$ and ${\bf R}(t)$ get the same direction implying a direct correlation between the respective components of the two parameters. It also implies 
 \be
 E(t)=k\frac{Q}{R^2(t)}.
  \label{efield2}
 \ee
 
 Treating the case with $Q=1/k$, with no changes in the course of time, we can write
 \be
  { E}(t)=\frac{ 1}{R^2(t)}
  \label{field_1}
  \ee
  \subsection{Geometry of space-time parameters}
 In the static condition ($v=0$), $\bf R$ is the linear distance between the source and the detector. Instead, in the present case, that distance changes continuously in such a way it follows a curve section rather than a linear segment. If the assertion can appear conflicting with the absolutely straight way of the signal propagation, it is useful to note that although the distance represents a linear parameter, its changes follow a quadratic law. Anyway, for simplifying the analytic computation, instead of the entity $\bf R$, we can consider the equivalent vector $\bf dR$ with the same absolute value of ${\bf R}$ but  originated in an ideal "space-time" point, evidently different from any position of the source trajectory:
 \be
 {\bf dR}(t)\equiv R(t)~\hat{\bf R}(t)=dR~\hat{\bf R}(t)
 \ee  
Fundamentally, defining $\bf dR$ as a vector, we just intend that the signal path can be described using  just two points separated by the length $dR$, or equivalently, we can say about $\bf d R$ as aligned on $\bf c$ direction.  So, we constrain an extremity on $\bf Y$, we can put the other one at ${\bf O_R}$, that is whichever the following relation is satisfied 
\be
{\bf dR}={\bf Y}-{\bf O_R}
\label{deltar}
\ee
As can be deduced from the fig.\ref{time_0} (until \ref{ctd2}), $\bf O_R$ can be also understood as the $locus$ of  the points at same temporal distance from $\bf Y$, and it is the surface of the sphere with radius $dR$ and centered at $\bf Y$.
Establishing the length $dR$ developed between two instant $t_0$ and $t_Y$, it is very intuitive and useful to develop $dR$ in terms of measurable parameter. That implies 
\be
dR=c(t_Y-t_0).
\label{cdt0}
\ee
Obviously, $t_0$ is the signal starting time and $t_Y$ is the detection instant.

Translating the definitions by eqs.\ref{deltar} on the trajectory direction, and using the eq.\ref{track} with the eq.\ref{cdt0}, we get
\be
x_Y-x_0=\frac{v_x}{c}dR=\beta dR
\label{deltax}
\ee
 with $\beta \equiv v_x/c$, and having defined
\be
x_0\equiv x(t_0)=v_xt_0~~~~~{\text and}~~~~~ x_Y\equiv x(t_Y)=v_xt_Y.
\label{x0xy}
\ee
 \subsection{ Correlation between signal and source paths }
\begin{figure}
\includegraphics[width=0.8\linewidth,keepaspectratio]{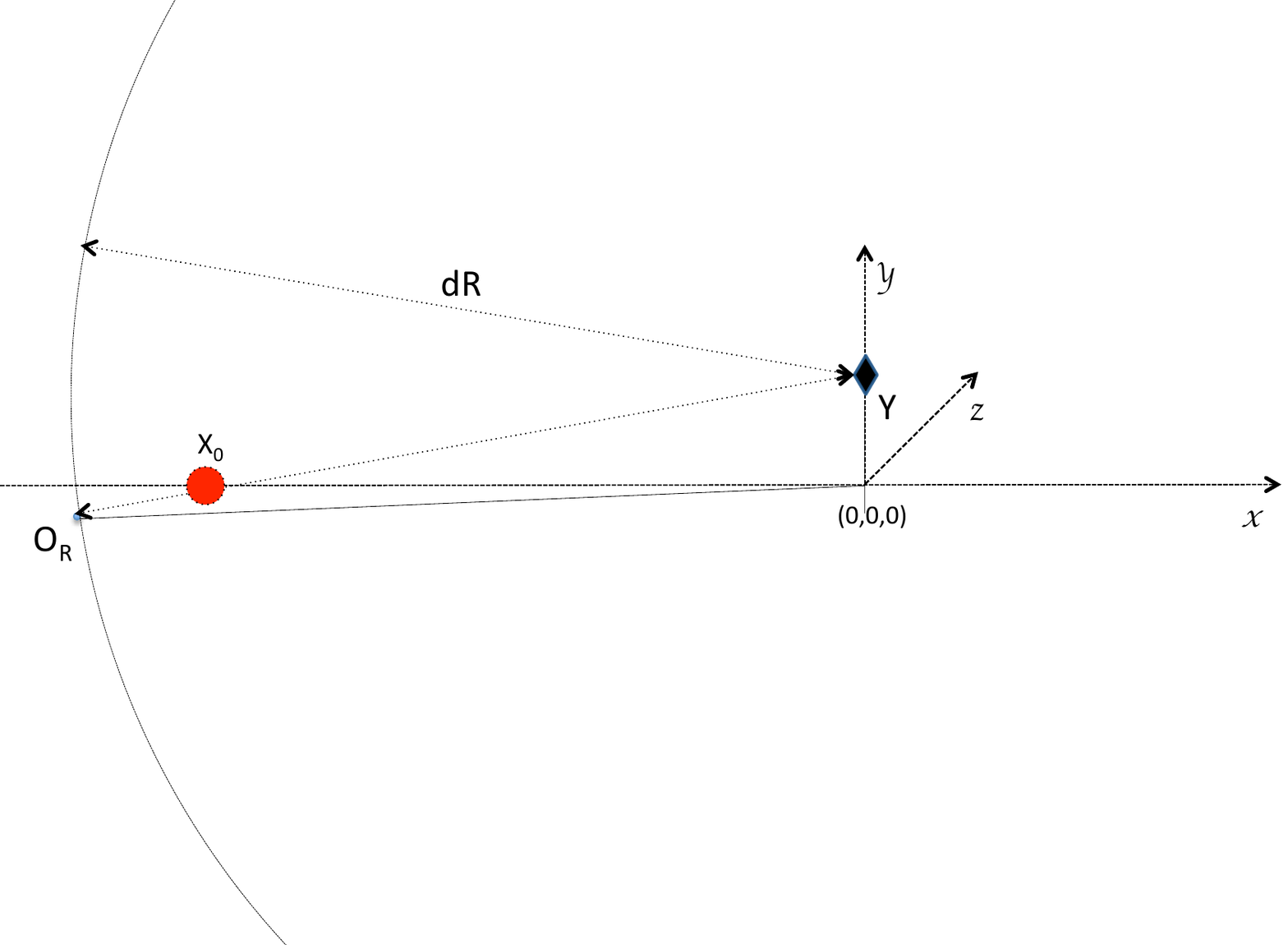}
\caption{
In the figure, the reference system with origin $(0,0,0)$, and the detector position on the $y$-axis are shown. The red full dot indicates the source position at instant $t=t_0$. The circle centered at ${\bf Y}$ represents the $x-y$ section of a sphere with radius $dR$. That quantity is deduced from eq.\ref{dr_x0}, adopting arbitrary units such that Y=2, $c=1$ and $\beta =v/c=0.75$. 
The quantity $\bf dR$ is also represented by the vectorial relation $\bf O_R-Y$ with module always greater than $\sqrt{x_0^2+Y^2}$, the distance between the charge and the detector at time $t_0$. It is important to underly that at $t=t0$, $dR$ is just a geometrical entity, but actually it is still null. Intuitively, it can be very useful interpreting the circle line as the $zero~locus$ for the quantity $\bf dR$.
\label{time_0} }
\end{figure}

 As resulting from the section \ref{expe}, determining  $\bf R$ is fundamental to evaluate the effects from an EM source. So, it is a primary matter to analyze the dependence of $R$, or the equivalent quantity $dR$, on absolute measurable parameters that are non-dependent on any specific reference system. Debating about time quantities, it results very difficult to obtain a clear and exhaustive graphic representation. First of all, we try to obtain it referring to a specific experimental configuration that includes the following numbers in arbitrary units (AU): $c=1$, $v_x=0.75~c$, and $Y=2$.
  In the fig.\ref{time_0}, the reference system and some of the previous defined quantities are shown, as the source position at instant $t_0$, and also the dimension $dR$, the quantity that will be covered between $t_0$ and $t_Y$ at $c$ velocity, represented by any radius of the sketched circle centered  at $\bf Y$. At beginning, it could seem a stretching to draw $dR$ before determine the relative source displacement. Nevertheless physically, given a temporal coordinate of the source, it is possible univocally to establish the relative length of the signal path, as following it will be shown.\\
 
\begin{figure}
\includegraphics[width=0.8\linewidth]{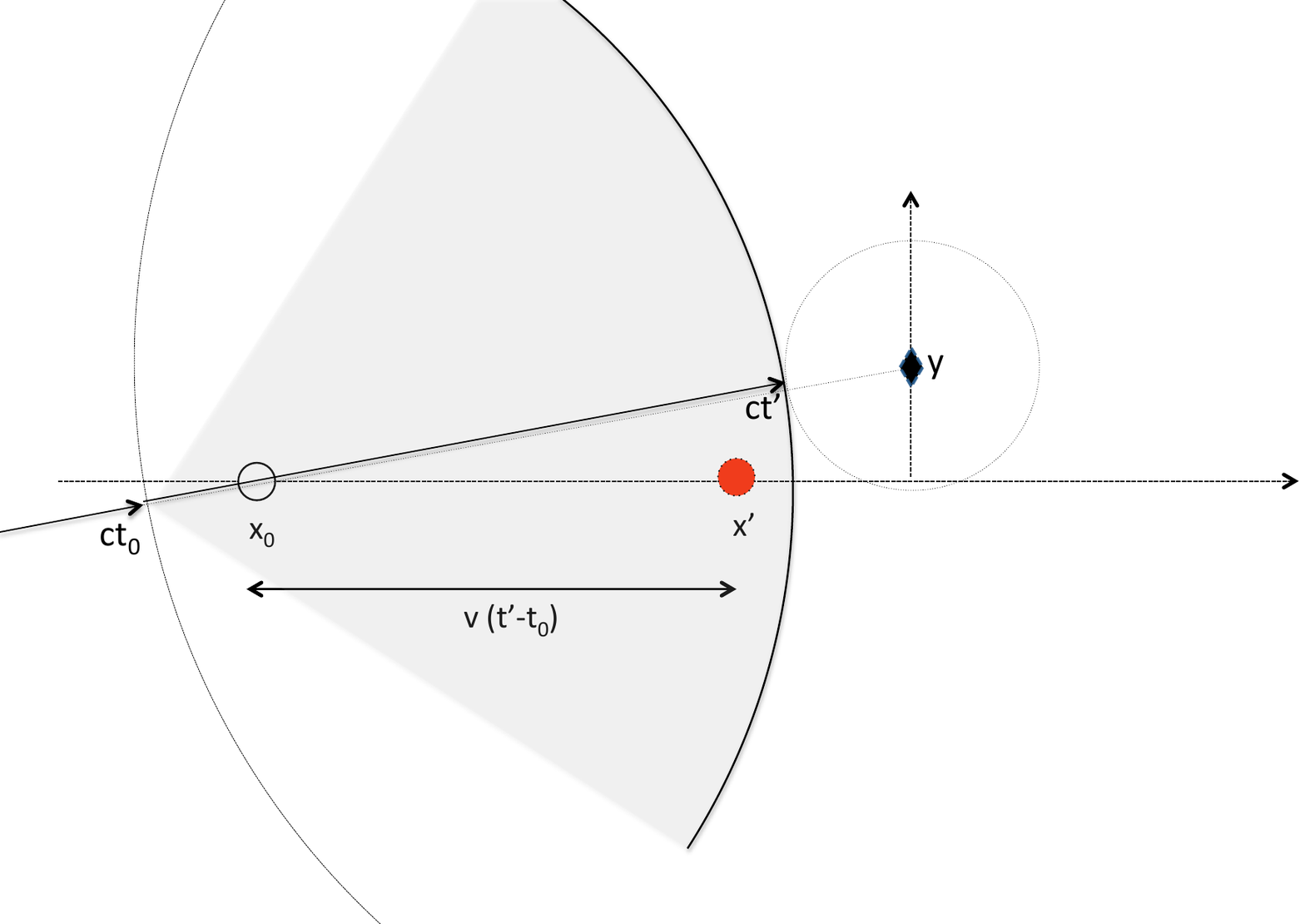}
\caption{
The plot presents the same previous configuration but at instant $t=t'$ following $t_0$ such a way the eq.\ref{tapex} holds. The traveling charge is at $x'$ point (red full dot). While on radial dimension the increase is $c(t'-t_0)$,  the source moves over $v_x(t'-t_0)$ on $x$-axis. Being the real signal on the surface of the shady sphere, it is clear that not yet signal is detected at $\bf Y$. From this point of view, the little circle represents the signal front-end whose radius is [$dR-c(t'-t_0)$].
\label{ctp1} }
\end{figure}

\begin{figure}
\includegraphics[width=0.8\linewidth]{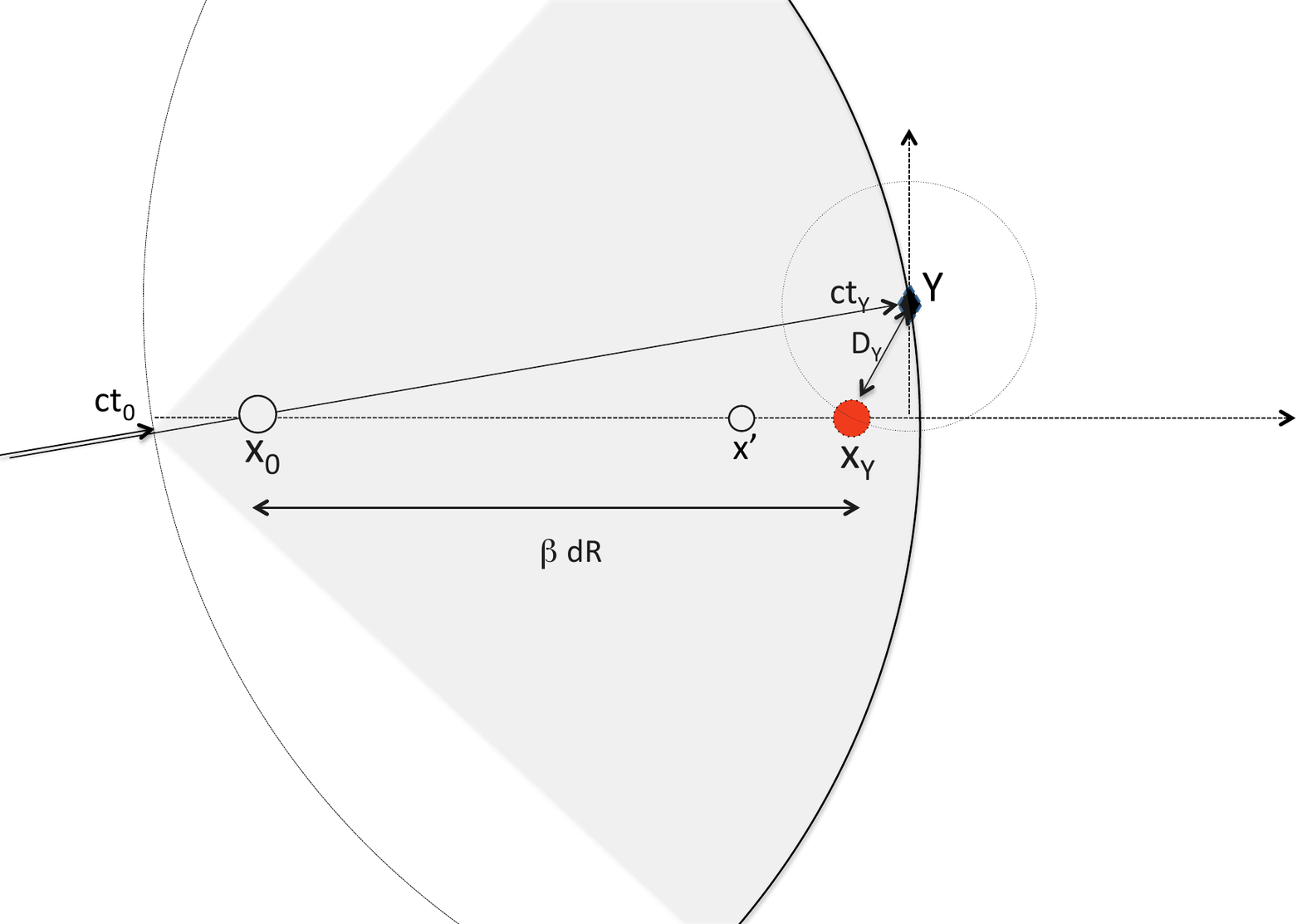}
\caption{The figure shows the source position $x_Y$ (red full dot) at time $t=t_Y$ such that the eq.\ref{errep} holds. With respect to the fig.\ref{time_0}, charged source displacement of $\beta dR=v_x(t_Y-t_0)$ occurs (see also the further figure), while with respect to the fig.\ref{ctp1}, a radius displacement of $c(x_Y-x')$ happens. At that instant, the detector gains $E=1/dR^2$ in terms of field intensity. Anyway, it must be clear that no signal will be detected since $t_0$  until $t_Y$, unless it would be generated at an instant preceding $t_0$. 
\label{cdt1} }
\end{figure}

\begin{figure}
\includegraphics[width=0.8\linewidth]{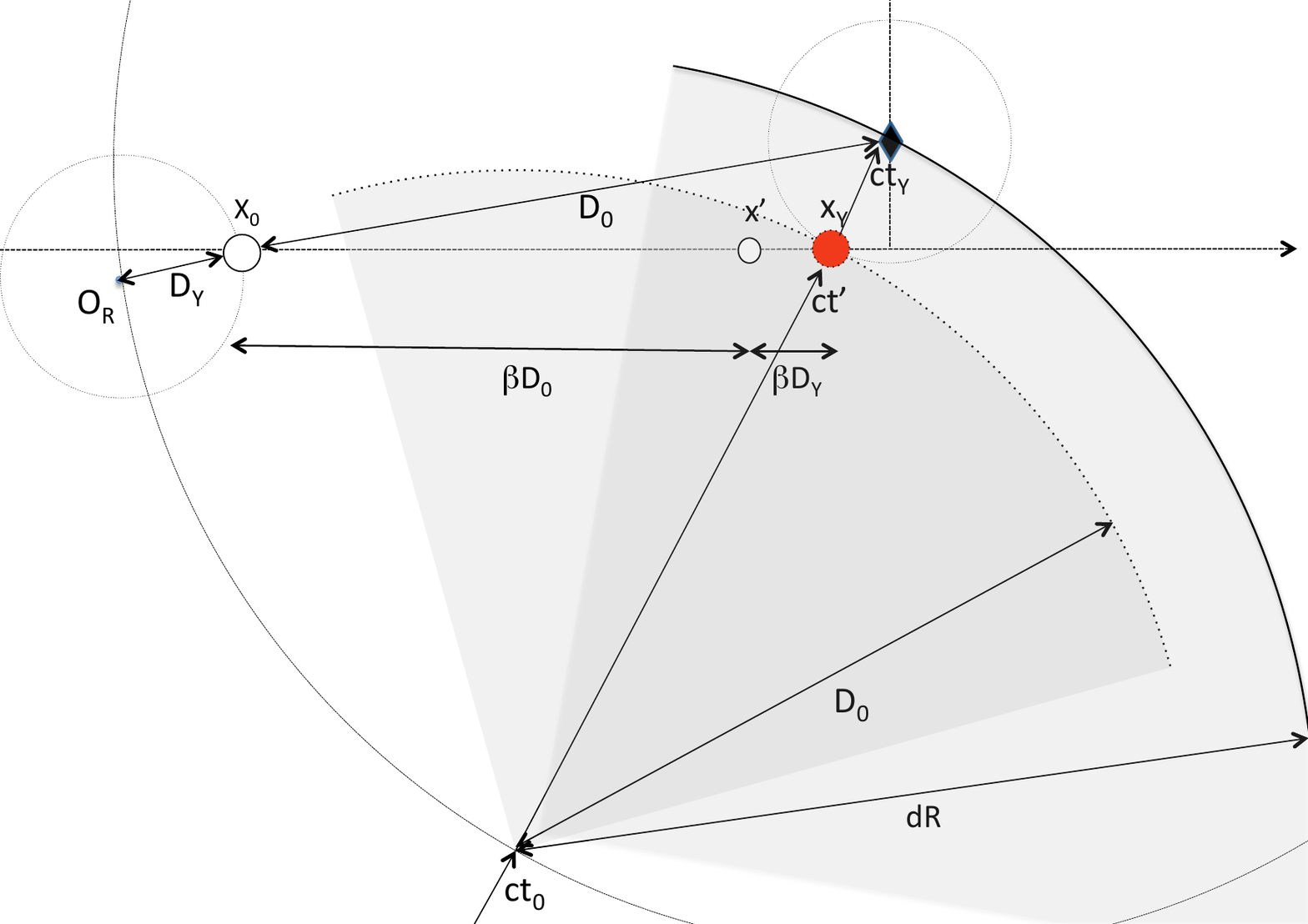}
\caption{
The figure describes the same case of the previous one, showing the growing signal on another radial direction, to demonstrate the equivalence with the previous configuration, seeing that the signal origin is on another point respect to fig.\ref{time_0} but on the surface of the same previous sphere with radius of $dR$ and centered at $\bf Y$. The figure is been reported also to complete the relative parameter description without overload the notation, and so better understand the relations reported in the text.
\label{ctd2} }
\end{figure}

Resuming the eq.\ref{deltar}, we deduce that it holds true for any $\bf O_R$ belonging to the sphere surface with radius $dR$. On this length, the EM signal travels at light speed $c$, so we can express $\bf dR$ as the difference between two aligned vectors ${\bf c}t_Y$ and ${\bf c}t_0$, (see figs.\ref{time_0}-\ref{ctd2}):
\be
{\bf c}~t_Y-{\bf c} ~t_0={\bf c} ~dt
\label{cdt}
\ee
being $dt=t_Y-t_0$, with $t_0$ and $t_Y$ as defined before.
 Although the time parameter is not an absolute quantity, we anyway recognize  a vector with the vertex at $\bf Y$, at time $t_Y$, and as described in fig.\ref{ctp1}, it is always possible to compose it with a temporal term $t'$ such that 
\be
ct_Y=ct'+\sqrt{x_Y^2+Y^2}
\label{errep}
\ee
 Looking at fig.\ref{cdt1}-\ref{ctd2},  noticing the geometrical characteristic of the quantity $c(t'-t_0)$, we also understand
\be
ct'=ct_0+ {\sqrt{x_0^2+Y^2}}
\label{tapex}
\ee
from which, together with the previous relation, we deduce
\be
cdt=dR=\sqrt{x_Y^2+Y^2}+\sqrt{x_0^2+Y^2}.
\ee
Using the last equation together with the eq.\ref{deltax}, after a few algebra, we get
\be
dR=2\frac{(D_0+\beta x_0)}{1-\beta^2}
\label{dr_x0}
\ee
or
\be
dR=2\frac{(D_Y-\beta x_Y)}{1-\beta^2}
\label{dr_x}
\ee
where
\be
D_0\equiv \sqrt{x_0^2+Y^2}~~~~\text{and}~~~~D_Y\equiv\sqrt{x_Y^2+Y^2}
\ee

The eqs.\ref{dr_x0} and \ref{dr_x} show how the variable $dR$ is predictable, consequently also the field intensity, just knowing  the source path that is exactly the space time correlation during the  displacements, QED.\\

From the experimental point of view, it results useful explicit the quantity $dR$ also in terms of $t'$ as well as $x'\equiv x(t')$. Using this last definition, the eqs.\ref{errep} and \ref{tapex}, the following relation is obtained
\be
dR=2\frac{\sqrt{(x')^2+Y^2(1-\beta^2)}}{1-\beta^2}.
\label{drp}
\ee
The last equation results very useful to calculate the maximum field value for each specific environment, being
\be
E_{max}=1/dR_{min}^2
\label{emax}
\ee
with
\be
dR_{min}\equiv dR(x'=0)=2\gamma Y
\label{dr_min}
\ee
where $\gamma$ is  $1/\sqrt{1-\beta^2}$.

\begin{figure}
\includegraphics[width=1.2\linewidth]{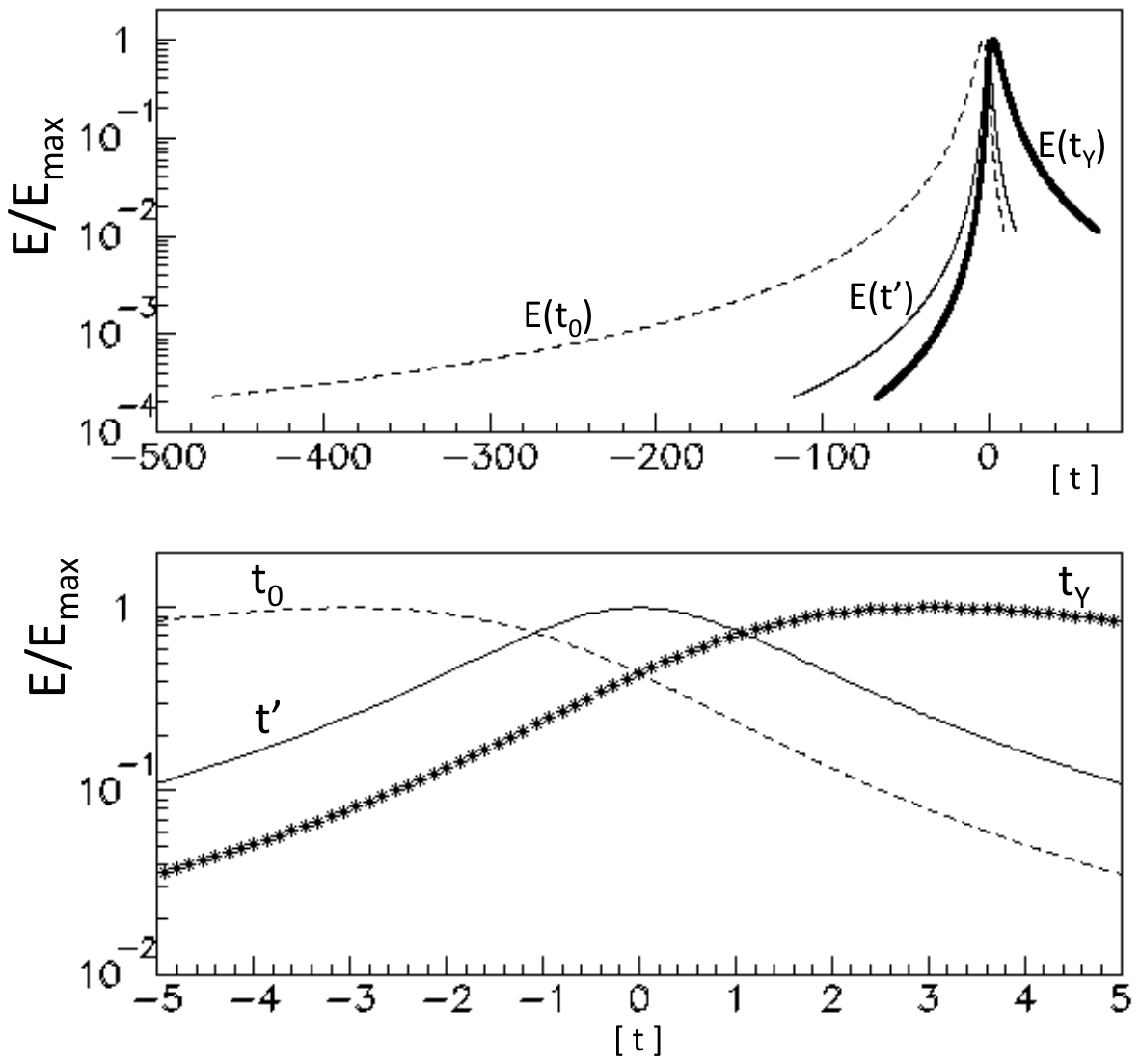}
\caption{
The graphics describe the behavior of the field intensity $E=(1/dR)^2$ detected at $Y$ as fixed before, normalized to the maximum value $E_{max}$. The experimental parameters are just that used before (see Example 1 in the text). The lower graphic is a zoom of the upper one. On the $x$-axis, the time units are reported. The star points describe $E(t_Y)$ as a function of the real current position of the field source, $x_Y$. The dotted line represents $E(t_0)$, that is assuming $t_0$ as free parameter. The continuos line describes $E(t')$ where $t'$ is defined by the eq.\ref{errep}. As it can be deduced from the eq.\ref{drp}, $dR$ and consequently $E(t')$ are symmetrical with respect to $x'=0$. (Often in laboratory measurements, that curve is used for evaluating the source tracking, but any experimental operator must be careful exchanging $x$ and $x'$ because they are correlated through \ref{errep} that is not a linear equation! )
 \label{three_t} }
\end{figure}
\begin{figure}
\includegraphics[width=1\linewidth]{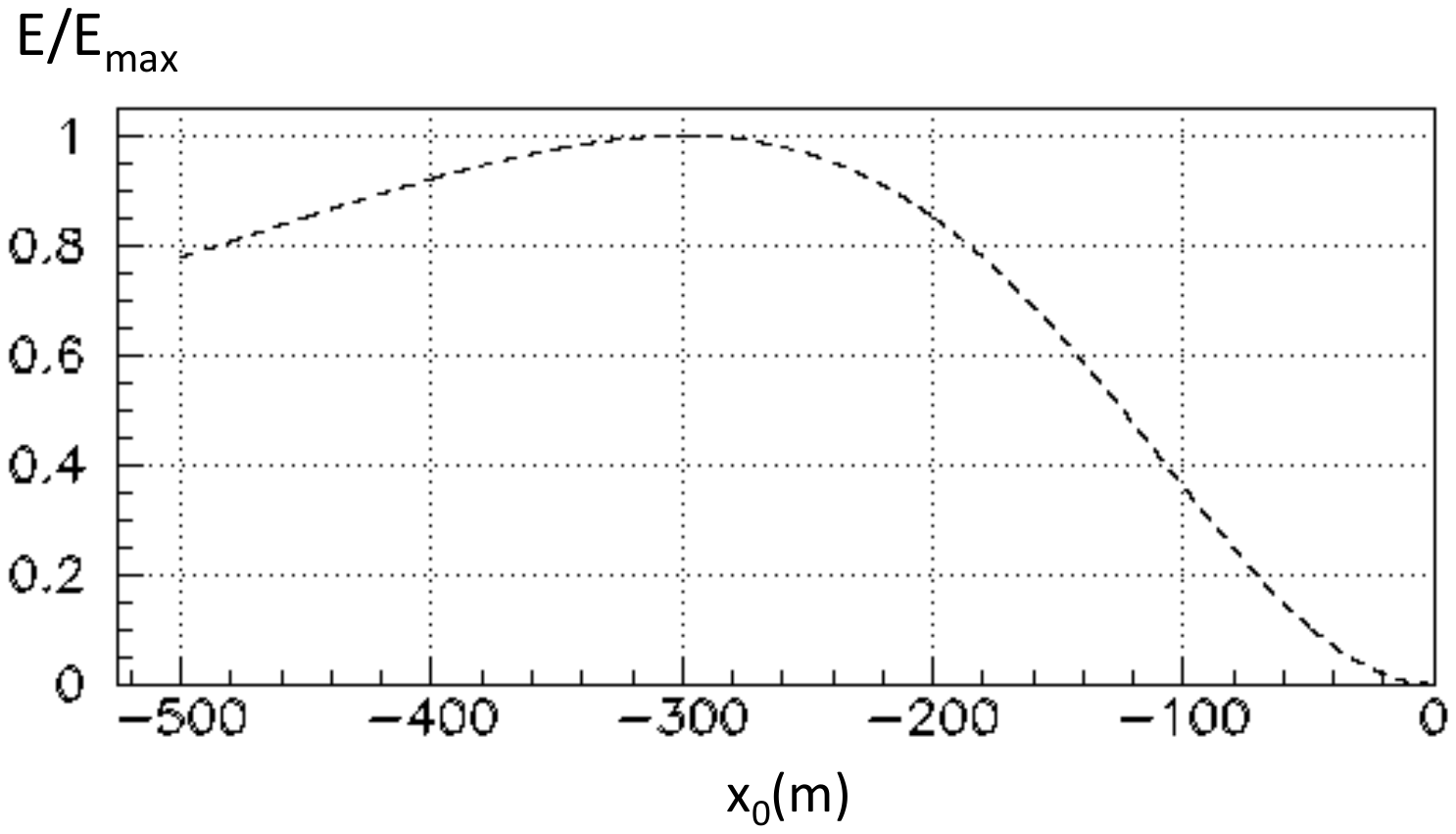}
\caption{The graphic shows the behavior of the normalized function $E(t_0)$ to its maximum value, in a experimental context with $\gamma=1000$ and $Y=30cm$ (see Example 2 in the text). The physical parameters are equal to the example contained in the  ref.\cite{guido}. As can be easily verified comparing the present graphic with the fig.2 of the cited article, the curve fits with that obtained using the LiŽnard-Wiechert formula.
 \label{guido1} }
\end{figure}
\section{Some specific examples}
\subsubsection{ } Searching for an explicit correlation between the intensity field and the source trace, in the previous section, we used a specific numerical example in order to explain and better represent the physical context. It is concerning an EM source releasing a radial signal at $c=1$ rate, and moving on $x$-axis at $v_x=0.75~c$. The fig.\ref{time_0} shows the environment at $t_0$ time. The successive figure contains the system evolution until $t'$ instant. The figures \ref{cdt1} and \ref{ctd2} report the same instant $t_Y$ that is the record time of the signal $E=1/dR^2$, with the corresponding source position $x_Y$. So regarding the graphic notation, they are complementary each other. In the fig.\ref{three_t}, the signal as function of the simultaneous charged position is reported. Three curves are shown, but it is very important to underline the deep difference between them, because while the dotted line (eq.\ref{drp}) and the continuos line (eq.\ref{dr_x0}) symbolize just the evolutions of reciprocal geometrical proportions, the third curve (eq.\ref{dr_x}), the star line, concerns physical quantities like the current time, the real source position and the real detected signal. It is also evident that three functions assume a completely different behavior with respect to the time variable, being not just shifted each  other. So from the experimental point of view, it is fundamental to fix the real source tracking as free parameter and not a generic variable depending on time, although with a linear law.

\begin{figure}
\includegraphics[width=1\linewidth]{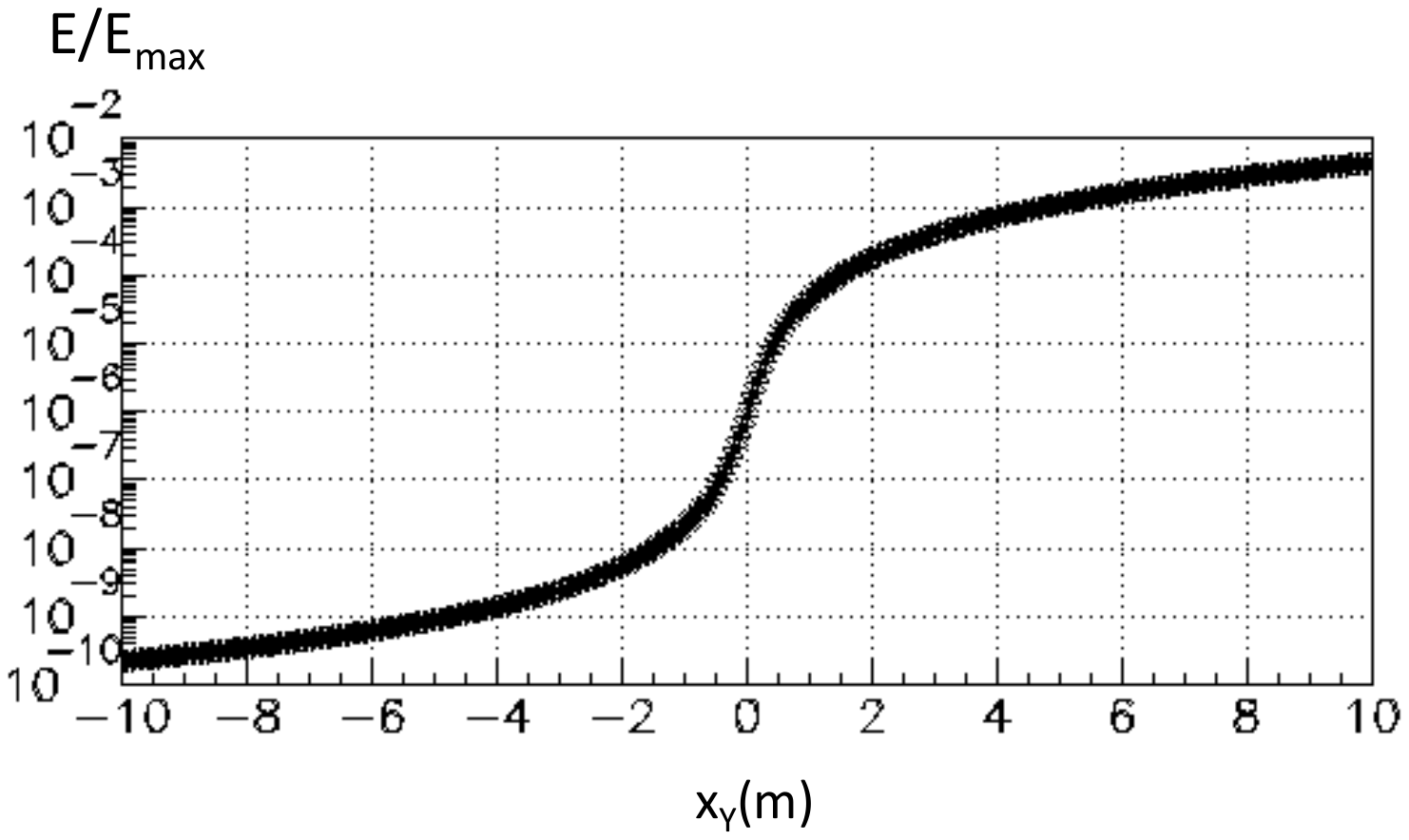}
\caption{The plot refers to the same configuration of the previous graph relative to the Example 2. Here the normalized signal $E$ is shown as function of the source position at the detecting time. The maximum value is not yet visible, since it will occur when $x_Y=300m$, out of the considered space interval.
 \label{guido2} }
\end{figure}

\subsubsection{ }To better say the extremely risky evaluating that kind of correlation, we show another example referring to the same experimental environment but with a different numerical set, getting it from ref.\cite{guido}. In the cited article, an experiment was performed to measure the propagation speed of Coulomb field, in order to interpret
some theoretical predictions that would induce to paradoxical facts. For instance, referring to the description of the EM field (fig. 2 in the ref.\cite{guido}) as a function of the source track, a paradox is shown since the maximum recorded signal would occur in an unphysical region, while the function assumes a negligible value passing nearer to the detector. In the our opinion, the incongruity can be solved interpreting  the variable $x'$ of the cited article like the variable named $x_0$ in the present paper. Effectively, in the fig.\ref{guido1}, the same quantity is shown as function of $x_0$, as deducing it from \ref{field_1}, \ref{deltar} and \ref{dr_x0}, using $c=$light speed, $Y=30$ cm and a source powered with a relativistic factor $\gamma = 1000$. Comparing that with the fig. 2 of the ref.\cite{guido}, we notice the maximum of the normalized field intensity occurring at the same abscissa value $-300 ~m$, further,  the two functions show the same behavior unless some light difference on the extreme parts where probably the effects of a different numerical approximation could mainly influence. As said before, to correlate the field intensity to the current source position, the variable $x_Y$ must be employed, and so the relation \ref{dr_x} must be used. In the fig.\ref{guido2}, relating to a typical laboratory dimension, the  normalized measurable quantity $E(t_Y)$ normalized to $E_{max}$ is reported for a source interval tracking of $[-10m,10m]$.

\begin{figure}
\includegraphics[width=1\linewidth]{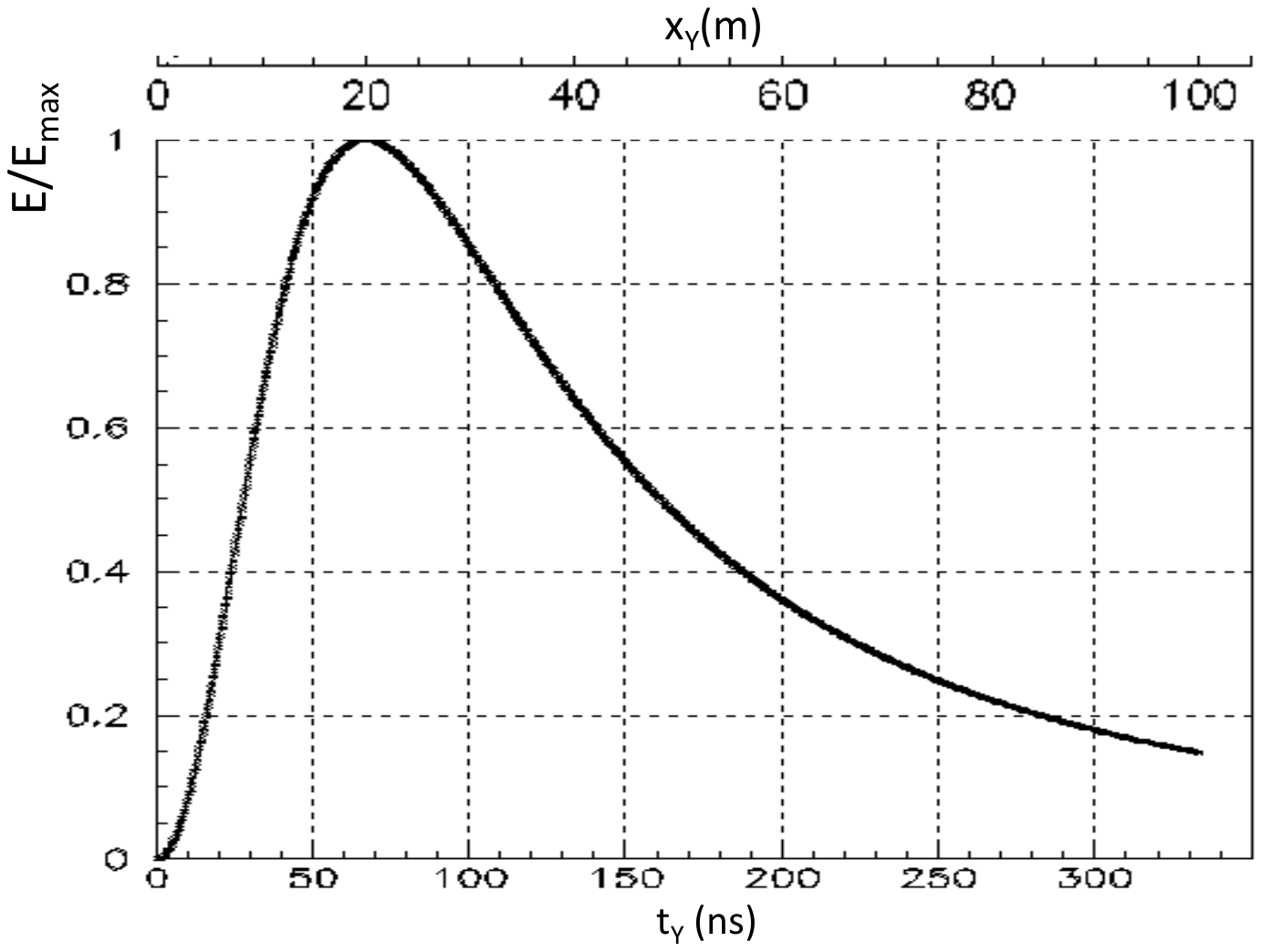}
\caption{The present graph is related to the Example 3 ($\gamma=400$). It shows the signal behavior while the charged source runs between $0$ and $100$ meters. The maximum value is well visible at $20$ meters, $67$ nanoseconds after the $zero$ reference time.
 \label{cern1} }
\end{figure}

\begin{figure}
\includegraphics[width=0.9\linewidth]{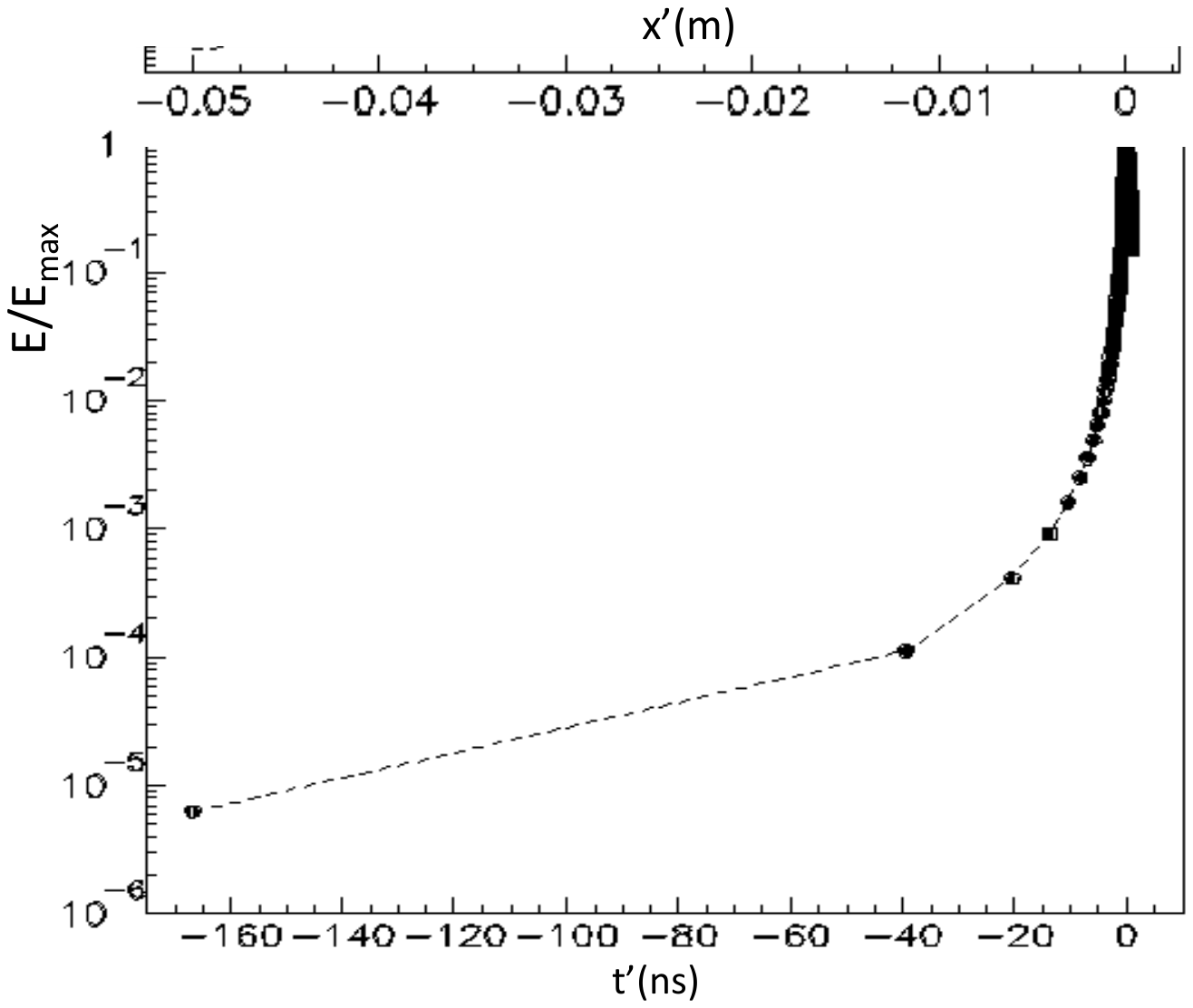}
\caption{As the previous one, the graph is related to the Example 3. Calculated by the eq.\ref{drp}, it shows the normalized signal as it appears as the function in terms of the time parameter $t'$ that is un intermediate value between $t_0$ and $t_Y$, the real detecting time of $E$. The function gets the maximum value at $t'=0$ and it results symmetrical around that point (see the zoom in the next figure), presenting a total difference relative to the previous figure, also in terms of numerical domain.  It records the maximum value twenty meters before compared with real position, causing a temporal mismatch of almost $67$ nanoseconds for the EM source traveling at ultra relativistic velocity. 
\label{cern2} }
\end{figure}

\begin{figure}
\includegraphics[width=0.9\linewidth]{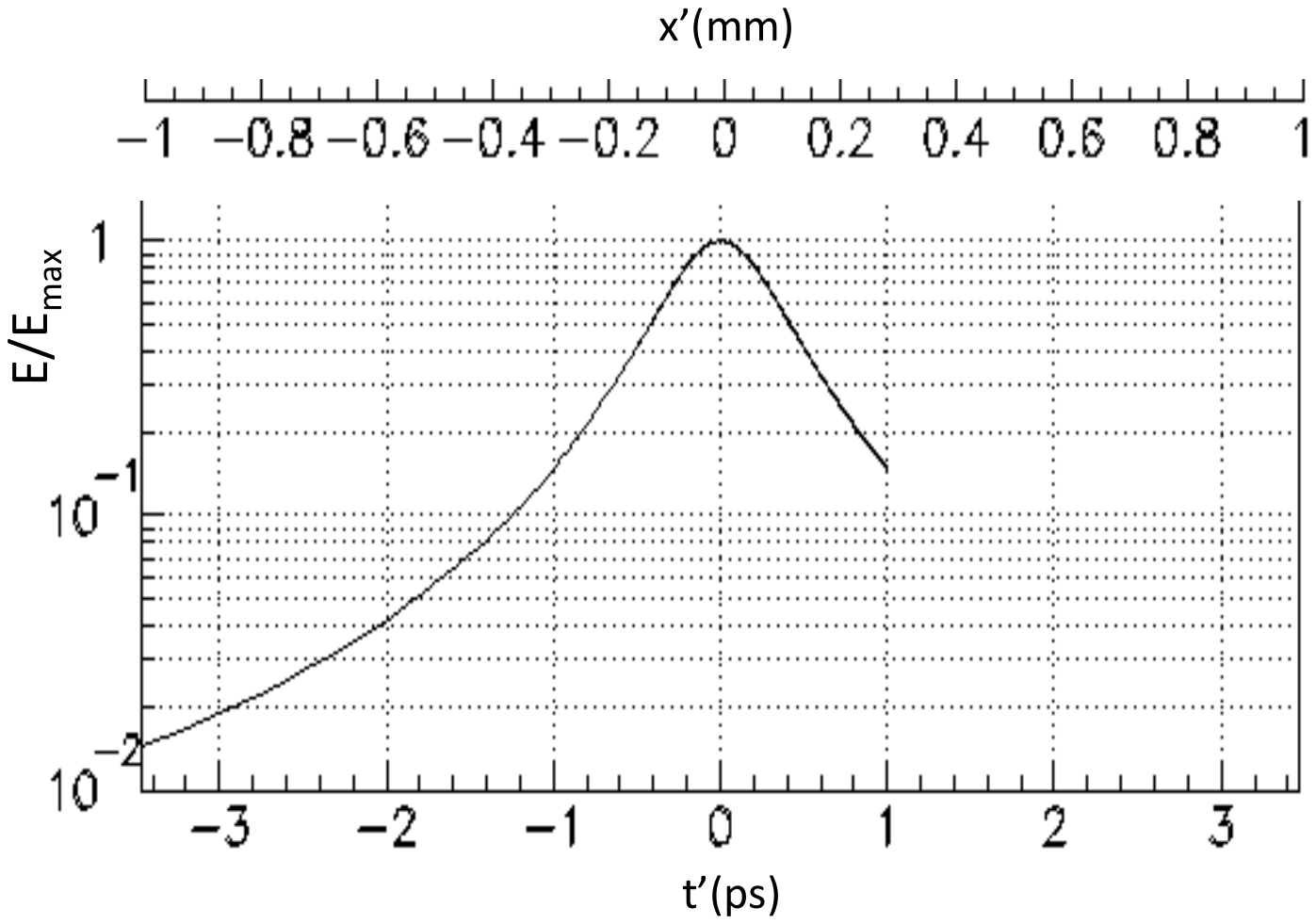}
\caption{The graph is a zoom of the previous fig.\ref{cern2} around the zero value both for $t'$ and $x'$.
\label{cern3} }
\end{figure}

\subsubsection{}
In this example a punctual relativistic source powered with $\gamma$ factor of $400$ is considered. The detector distance is $Y=5~cm$ from the origin of the reference system. Using a space interval $dx=[0,100m]$, calculating the radial quantity $dR$ as the function of the current source position, that is using eq.\ref{dr_x}, we obtain the curve  as in the fig.\ref{cern1} where we can see the maximum occurring at $20m$, very far from the point at minimum distance from the detector. In temporal terms, it corresponds to $67ns$ about. Using a metrics described by the $x'$ parameter that evidently does not represent  the trajectory, the same signal evolves in a space interval $[-50mm,0.3mm]$, and it shows the maximum at $x'=0$, as shown in fig.\ref{cern2}. In the fig.\ref{cern3}, we can see a zoom of the same function.

\section{Discussion and conclusions}
In the present report, a very simplified scientific context has been treated, aiming to increase as much as possible the awareness about the fundamental meaning and the physical weight of some parameters, evaluating the  effects of an EM moving source. Using an elementary formalism, the correlation between the signal and geometrical parameters has been obtained as well as their time-dependence. Being the time representation the principal hindrance, that has been avoided developing the time parameter $t$ on any linear radial dimension $\bf c$$t$,  centered at the detector position, while the corresponding source position is $\bf v$$t$. 

Three time parameters are been identified, correlated to the same signal $E$ :\\

$t_0$ representing the instant the signal is at $dR$ distance from the detector,\\

 $t'$, when the same vector  crosses the $x$-axis, and\\
 
$t_Y$, time detection of $E=1/dR^2$, being $dR=c(t_Y-t_0)$, and the source position $x_Y=v_xt_Y$.\\

 To be faithful to the main spirit of the job, i.e. to make the physical parameters intuitively understanding, we can associate them to everyday concepts. For instance, let's image to observe a highway from a bridge, and to take a picture (by night, the example is more pertinent) of a running car. In this context, $t_0$ represents the camera opening instant. The light will reach the camera device at $t_Y$, when the car is at $x_Y$, paradoxically a position that we can only foresee but not see. The time $t'$ regards the position $x'$ that we illusory charge to the signal recording time $t_Y$.\\
 
 Coming back to the mathematical discussion, we note that although their specific smart form, the main relations steadily hold their non-linear nature, as clearly the equations \ref{errep}-\ref{drp} show, so implying an high level of criticality performing discretional opinions or some numerical approximations. Being consequently well aware that no practical example could be completely exhaustive, nevertheless, three typical experimental settings are been reported corresponding to several $\gamma$ factors, $\gamma_1 =1.512$, $\gamma_2=1000$, and $\gamma_3=400$. The first one aiming just to explicit and to graphically represent the physical parameters, the second one trying to solve some paradoxes as requested in the \cite{guido}, and the third one aiming to demonstrate how a wrong interpretation of the time parameters could determine dangerous misleadings about the signal evaluation or the real position of the field source, as well as about the true velocity of the same source. 
 
 In summary, in this kind of experimental setting, to avoid misleading physical predictions, it results very useful taking into the account just one between the formulae \ref{dr_x0}, \ref{dr_x} and \ref{drp}, but being solidly aware about the choice of the temporal free parameter.\\

{\large\bf Acknowledgements }\\

I thank Giorgio Fornetti, insightful referee for every discussion.



\begin{thebibliography}{99}
\bibitem{fey0}"...Whenever you see  a sweeping statement that a tremendous amount can come from a very small number of assumptions, you always find that is false. There are usually a large number of implied assumptions that are far from obvious if you think about them sufficiently carefully." \cite{fey1}
\end{thebibliography}
 \end{document}